# Privacy issues on biometric systems[1]

Marcos Faundez-Zanuy
Escola Universitaria Politècnica de Mataró
Avda. Puig i Cadafalch 101-111
08303 MATARO (BARCELONA) SPAIN
E-mail: faundez@eupmt.es   http:www.eupmt.es/veu

**ABSTRACT**
In the XXIth century there is a strong interest on privacy issues. Technology permits obtaining personal information without individuals' consent, computers make it feasible to share and process this information, and this can bring about damaging implications. In some sense, biometric information is personal information, so it is important to be conscious about what is true and what is false when some people claim that biometrics is an attempt to individuals' privacy. In this paper, key points related to this matter are dealt with.

**Introduction**
When evaluating a biometric system, several questions must be addressed, including the vulnerability of different systems by fraudulent users [1], the system performance and how to improve it [2], and the privacy issues of the biometric system. This paper is related to this last topic, because several misunderstandings are established around it.

**Personal data**
Technological advances let to store, gather and compare a wide range of information on people. Using identifiers such as name, address, passport or social security number, institutions can search databases for individuals' information. This information can be related to salary, employment, sexual preferences, religion, consumption habits, medical history, etc. Though in most of the scenarios there should be no problem, there is a potential risk. Let us think, for instance, in sharing medical information. Obviously, in case of emergency, this sharing between hospitals would be beneficial. On the contrary, if this information is transferred to a personal insurance company or a prospective employer, the insurance or the job application can be denied. Due to these facts, there is a controversy about privacy issues, and biometric systems are subject to this. Obviously, special control must be taken to ensure that data cannot be used beyond the purposes it was originally collected for, but this cannot always be assured. An example of this situation is [3] the Social Security Number (SSN) in the United States. Originated in 1936, its sole purpose was to facilitate recordkeeping for determining the amount of Social Security taxes to credit to each contributor's account. The original SSN cards contained the legend "Not for identification". By 1961, the Internal Revenue Service (IRS) began using the SSN for tax identification purposes. Nowadays, "everything from credit to employment to insurance to many states' drivers licenses requires a SSN". From "not for identification", the SSN has become virtual mandatory identification.

According to [4], personal data can be any data that enables a physical person to be identified, such as his name, telephone number or photograph. National laws concerning data protection demand suitable practices, like the duty to handle data properly and safety, and to use personal data with legitimate and explicit purposes. Strict rules are applied to sensible data: those referring to ethnic or race origin, ideology, religious believes, syndicate affiliation, health or sexual life. As a general rule, these data cannot be processed. However, there are exceptions for concrete situations.

While biometric data tries to characterize us as individuals different from each other, there is another kind of data that can characterize our consumption habits. For instance, there are loyalty cards, especially in supermarkets. The advantage for the customer is that he gets discounts and points that can change for prizes. The advantage for the company is that he can rationalize the products' distribution along the supermarket, and take better care of their prices' politics, stock, etc. Consumers are not aware about the risk of permitting that the supermarket stores this information, although it is quite evident. Other companies would be willing to pay a lot of money for information about which kind of products is purchased by a given person (tobacco, alcohol, drugs, etc.), which kind of videos are hired, how much money spends each month, etc. Doubtless, consumption habits are more personal and private than our face, fingerprint or voice, and special care must be taken for this information, although the offers of the companies seem quite tentative.

---

[1] This work has been supported by FEDER and CICYT, TIC-2003-08382-C05-02

**Three important points to make clear**
Some misunderstandings that should be made clear are the following:
1. The first important aspect is to be aware that it is the application of biometrics, not the technology itself, which defines its relation to privacy [5]. For instance, a fingerprint can be used to protect personal information saved in a computer, which cannot be unlocked without the genuine user's finger. On the other hand, the fingerprint from the database can be extracted without user's consent, and compared against a latent fingerprint found on a crime scene, in order to check if this person is guilty or not. Thus, the same technology (fingerprint recognition) can violate or protect privacy depending on its use.
2. The second aspect is that biometric systems do not store the fingerprint, iris, face or hand images, nor the speech file. Except for forensic applications, biometric systems store a small file derived from the distinctive features of a user's biometric data, named template. For each biometric technology several templates are possible, being incompatible to each other. For instance, the templates extracted and used by the U.are.U fingerprint recognition engine of digital persona [6] cannot be introduced in the Precise Biometrics fingerprint recognition engine [7] to obtain a successful result. It is important to point out that the template extraction procedure is irreversible, which means that it is not possible to re-generate the fingerprint, speech, iris, etc., beginning with the template. Thus, an important advantage of the lack of biometric standards is the impossibility (in most cases) to transfer information between systems designed by different vendors, due to incompatibilities. If the same technology and algorithm were used in several different scenarios, one possibility to reduce potential privacy risks, would be, for instance, to use different fingers for work and home usages.
3. While the outcome of a search using a social security number is a perfect match (if it is found), independently on the number of people in the database, the biometric matching yields a probability or distance measure, but never an exact match. This is due to the inherent variability on biometric data. For several biometric technologies, like hand-geometry, the identification procedure cannot be done over a large database, due to the lack of enough discriminative capacity between individuals. Thus, it is difficult to implement a fully automatic system, because perfect matches are not possible.

**DNA: a special biometric identifier**
Main automatic biometric systems (face, speech, fingerprint, hand-geometry, etc.) are based on the same information that it is available for a human being. Thus, in general, they are not acquiring a deeper knowledge than with a "human recognizer" using a conventional method (inked fingerprint, photo, speech recording, etc.). There is no additional risk on adopting an automatic biometric system, when compared to a traditional one. However, there is a strong exception to this statement: DNA identification.

DNA matching technology is far from being a cheap, automatic and fast identification method. Nevertheless, it is expected that in the future, people identification from naturally-occurring oils on the skin will be possible, and it will replace or be competitive with the actual state-of-the-art biometric systems (the highest possible accuracy is achieved through DNA identification). In this case, it must be taken into account that DNA contains information about susceptibilities of a person to certain diseases. Thus, the abuse of genetic code information may result in discrimination, for example, for job applicants, medical and drive insurance, etc.

A similar problem may exist for retina and iris recognition. An expert can determine that a patient suffers from diabetes, arteriosclerosis, hypertension, etc., from examining the retina or iris. However, if the automated biometric system scans the iris or retina and converts it to a template, without attempting to extract health information, the relation between the template and patient's health cannot be established neither at the present moment nor in the future, and the privacy will be kept.

**Particular population cases**
For any biometric identifier, there is a portion of population for which it is possible to extract relevant information about their health, with similar implications to the ones described in previous section. For example, speech disorders, hair or skin color problems, etc. However, it can be argued that this same information can be obtained when the identification or communication is established between two human beings, so in this case biometrics does not imply any privacy violation against counterpart systems without biometric systems. On the other hand, other situations exist where some information can be theoretically inferred from the biometric scanning. An important question is what exactly is disclosed when biometric scanning is used. In some cases, additional information not related to identification might be obtained. For instance [3,pp.393] presents a list of these cases that includes:
- Some studies suggesting that fingerprints and finger images may disclose medical information about a person (chromosomal disorders such as Down syndrome, Turner Syndrome and Klinefelter syndrome, and non-chromosomal disorders, such as chronic, intestinal pseudo-obstruction, leukemia, breast cancer and Rubella syndrome).
- Several researchers reporting a link between fingerprints and homosexuality.

In [8,p.46] there is a set of references about statistical correlation between malformed fingers and certain genetic disorders.

Nevertheless, these hypotheses have not been proved nor agreed by the scientific community, and in case they would be true, perhaps it should not be a big problem. Let's think of handwritten texts and the possibility to extract relevant information about personality, state of mind, etc., although its validity is generally accepted, nobody fears to show his personal calligraphy nor destroys his wasted personal written documents in order to avoid that they could be acquired by third parties.

While the relationship between genetic disorders and fingerprints may be possible, it is hard to believe that a fingerprint, which is fully formed at about seven months of fetus development and do not change throughout the life of an individual [8,p.24], could be correlated with sexual preferences that can vary, or diseases that can appear and disappear during our lives.

Whichever the case, the population percentage that can suffer this drawback is probably small, and this inconvenience can be overcome in one of these ways:
- Using a multimodal biometric system, where the user can freely decide between several biometric identifiers, and reject the system that he considers that it may reveal private information.
- Assuring that the biometric system does not collect any raw biometric data, and it just stores and processes a template extracted from the biometric data.

**CONCLUSIONS**
As a conclusion, we can state that information is the key factor. We must be aware of when, where and to whom we are giving our biometric information, how and where it is stored, and which the purpose is. There are no added risks when compared to other identification methods (passport, cards, etc), and the risk is related to the technology use, rather than the technology itself.

Probably a common sense rule to avoid risks would be enough. This is similar to what happens with VISA, American Express, etc., credit cards payment. They are very useful and convenient for us, but there exist a potential risk that they will be used by third parties for a different use from the one we thought. However, we do not renounce to use them, perhaps because the risk of not using it is even worse (to bring with us a lot of cash with the consequent risk to be stolen, the change can be a fake banknote, etc). All we must take into account is to be aware of their use.